\documentclass[conference]{IEEEtran}
\IEEEoverridecommandlockouts

\usepackage{cite}
\ifCLASSINFOpdf
\else
\fi
\usepackage{amsmath}
\makeatletter
\newcommand{\pushright}[1]{\ifmeasuring@#1\else\omit\hfill$\displaystyle#1$\fi\ignorespaces}
\newcommand{\pushleft}[1]{\ifmeasuring@#1\else\omit$\displaystyle#1$\hfill\fi\ignorespaces}
\makeatother

\usepackage{scalerel}
\usepackage{tikz}
\usetikzlibrary{svg.path}
\usepackage{soul}

\definecolor{orcidlogocol}{HTML}{A6CE39}
\tikzset{
  orcidlogo/.pic={
    \fill[orcidlogocol] svg{M256,128c0,70.7-57.3,128-128,128C57.3,256,0,198.7,0,128C0,57.3,57.3,0,128,0C198.7,0,256,57.3,256,128z};
    \fill[white] svg{M86.3,186.2H70.9V79.1h15.4v48.4V186.2z}
                 svg{M108.9,79.1h41.6c39.6,0,57,28.3,57,53.6c0,27.5-21.5,53.6-56.8,53.6h-41.8V79.1z M124.3,172.4h24.5c34.9,0,42.9-26.5,42.9-39.7c0-21.5-13.7-39.7-43.7-39.7h-23.7V172.4z}
                 svg{M88.7,56.8c0,5.5-4.5,10.1-10.1,10.1c-5.6,0-10.1-4.6-10.1-10.1c0-5.6,4.5-10.1,10.1-10.1C84.2,46.7,88.7,51.3,88.7,56.8z};
  }
}

\newcommand\orcidicon[1]{\href{https://orcid.org/#1}{\mbox{\scalerel*{
\begin{tikzpicture}[yscale=-1,transform shape]
\pic{orcidlogo};
\end{tikzpicture}
}{|}}}}

\PassOptionsToPackage{bookmarks={false}}{hyperref}
\usepackage{hyperref}

\usepackage{graphicx}
\usepackage{url}
\begin{document}
\title{Optimal Participation of Price-Maker Battery Energy Storage Systems in Energy, Reserve and Pay as Performance Regulation Markets}

\author{
\IEEEauthorblockN{Reza Khalilisenobari,~\IEEEmembership{Student Member,~IEEE,} and Meng Wu,~\IEEEmembership{Member,~IEEE}}
\IEEEauthorblockA{School of Electrical, Computer and Energy Engineering, Arizona State University\\
Tempe, Arizona 85287--5706\\
Email: rezakhalili@asu.edu - mwu@asu.edu}

\thanks{This work was supported by the Power Systems Engineering Research Center.}
}


%Telephone: (800) 555--1212, Fax: (888) 555--1212}


\IEEEpubid{\makebox[\columnwidth]{978-1-7281-0407-2/19/\$31.00~\copyright2019 IEEE \hfill} \hspace{\columnsep}\makebox[\columnwidth]{ }}
\maketitle

\begin{abstract}
Motivated by the need of assessing the optimal allocation of battery energy storage services across various markets and the corresponding impact on market operations, an optimization framework is proposed in this work to coordinate the operation of an independent utility-scale price-maker battery energy storage system (BESS) in the energy, spinning reserve and performance-based regulation markets. The entire problem is formulated as a bi-level optimization process, where the structure of all markets is modeled considering the joint operation limits. The strategic bidding behavior of a price-maker BESS in a pay as performance regulation market is investigated. Additionally, a specific approach is introduced for modeling automatic generation control (AGC) signals in the optimization. Although the formulated problem is non-linear, it is converted to mixed-integer linear programming (MILP) to find the optimum solution. The proposed framework is evaluated using test case scenarios created from real-world market data. Case study results show the impact of BESS's price-making behavior on the joint operation of energy, reserve, and regulation markets.  
\end{abstract}
\begin{IEEEkeywords}
Battery energy storage system (BESS), bidding strategy, price-maker, performance-based regulation market, bi-level optimization, mixed integer linear programming
\end{IEEEkeywords}

% \fi
\IEEEpeerreviewmaketitle

\section{Introduction}
In recent years, US power systems saw a growing integration of utility-scale battery energy storage systems (BESSs). With the capability of providing both energy arbitrage and fast ramping services, BESSs offer promising solutions to improve system flexibility and address renewable intermittency and uncertainty\cite{ref1}. BESSs' capability of multiple service provision is recognized in the smart grid roadmap of National Institute of Standards and Technology (NIST)\cite{ref2}. To further encourage BESS integration, US Federal Energy Regulatory Commission (FERC) issued a series of orders (784, 755, and 841), allowing BESSs to participate in various markets\cite{ref3,ref4,federalorder}.\par

The growing BESS integration has inspired researchers to evaluate the impact of BESS resources on system operation and address potential problems caused by BESS integration. Among existing research directions, optimal scheduling of merchant BESSs and the impact of BESSs' profit maximization activities on market operations are of great interest to power system society, especially when the BESS is large enough to perform as a price-maker. The work presented in this paper falls within this research direction.\par

Existing literature along the above research direction falls into two categories. The first category models BESS as a price-taker in various markets\cite{ref18,ref15,ref16,ref19,ref14,ref17}. An optimization framework for coordinating the participation of price-taker BESS in day-ahead (DA) an real-time (RT) energy markets is proposed in \cite{ref18}. Reference\cite{ref15} solves the profit maximization problem for a BESS that participates in energy, reserve, and performance-based regulation markets. Optimal charging/discharging schedule based on the battery's aging cost is achieved through an online optimal control algorithm for the operation of a BESS in regulation only \cite{ref16} or regulation and energy markets\cite{ref19}. Uncertainty in market price prediction and amount of energy deployment for the BESS strategic biding problem are handled by robust optimization approach in\cite{ref14}, and this work is expanded by adding aging cost to it in\cite{ref17}.\par

This paper falls into the second category of existing literature, where the BESS is modeled as a price-maker due to its size and specific operation capabilities\cite{ref25,ref22,ref23,ref24}. These papers not only solve the bidding strategy problem but also propose a framework for analyzing the impact of utility-scale BESS on various electricity markets. This impact is crucial for investigating the current expansion of utility-scale BESSs. Performance of various market mechanisms in existence of a utility-scale BESS is evaluated in \cite{ref25}. Coordination problem of a price-maker BESS in the DA energy market is addressed in \cite{ref22}. Although this work does not model ancillary services markets that are usually more appealing for a BESS, it performs a comprehensive analysis on the markets' outcome and storage profit in different conditions. Participation of BESS as a price-maker entity in DA energy and reserve markets along with RT balancing market is discussed in \cite{ref23} and\cite{ref24}. Considering DA reserve and RT balancing markets in these papers makes their analysis more accurate and closer the actual opportunities that price-maker BESSs can have. However, these papers do not model the frequency regulation market, therefore ignore BESSs' profit for regulation service provision. As an increasing number of BESSs are motivated by FERC order 755 for participating in the regulation markets, this exclusion may lead to inaccurate assessment on BESSs' stacked services and total profit.\par

In this paper, an optimization framework is proposed to coordinate the operation of a utility-scale price-maker BESS in the energy, spinning reserve, and regulation markets. The main contributions of this paper are as follows:

\begin{itemize}
    \item The bidding and operation problem for a price-maker BESS in energy, spinning reserve, and performance-based regulation markets is formulated as a bi-level optimization problem while considering operational details of BESS  and structural elements of each market.
    \item BESSs's fast ramping capability and accurate regulation signal tracking ability enable them to not only gain more profit from the pay as performance regulation markets but also become price makers and impact the market outcomes. Thus, it is crucial to consider the participation of price-maker BESS in the performance-based regulation market. To the best of our knowledge, this problem has not been addressed before.
    \item Deployment of automatic generation control (AGC) signals in a bi-level problem and coordination between energy, reserve and regulation markets are handled through a particular choice of market clearing intervals. 
    \item A realistic procedure for generating synthetic price bids and load data based on real-world market data is proposed in this work. Case studies for analyzing the operation scheduling of BESS and the impact of BESS scheduling on various markets are performed.
\end{itemize}

\par The remainder of the paper is organized as follows. Section II introduces the market structures adopted in this paper and proposes the price-maker BESS profit maximization model. Simulation procedure and different case study results along with the discussions on them are presented in section III. Finally, section IV draws conclusions and presents future research directions.

\section{Methodology and Formulation}
Since price-maker BESSs can impact market clearing outcomes, the strategic biding problem of BESSs is formulated as a bi-level optimization problem. In the upper-level (UL) problem, the BESS owner maximizes the revenue from participating in energy, spinning reserve, and regulation markets while considering its operating limits. The market clearing prices (MCPs) and scheduled power inputs/outputs of BESS in each market are obtained from the lower-level (LL) problem. The LL problem represents the joint optimization of energy, reserve, and regulation markets. This LL problem simulates the RT joint market clearing process of the independent system operators (ISOs). Note that in the strategic biding problem of a price-taker BESS, LL problem does not need to be modeled as BESS cannot affect markets, and prices are parameters for its profit maximization problem. Formulations of the UL and LL problems are presented in detail in separate subsections after depicting an overview of the joint market modeling.

\subsection{The Market Structures}
This section describes the market structures for the RT energy, spinning reserve, and regulation markets that are modeled in the proposed framework. 

The RT energy market is modeled using a simplified approach. This model allows BESSs to submit supply or demand bids in order to gain profit from arbitraging energy between different time intervals or various markets. Other market participants (such as conventional generators) submit supply bids to the RT energy market to fulfill system net demand in each time interval.\par

The RT spinning reserve market is an upward-only reserve market. Although it is not usual to consider a downward reserve market, it is similar to the upward reserve and can be handled by the proposed framework through minor adjustments. Market participants (including BESSs) are compensated by the MCP for their reserved capacities. Deployment of spinning reserve product is not modeled in this paper since the reserve deployment is called in the contingency situations, which falls out of the scope of this work.\par

The RT pay as performance frequency regulation market modeled in this paper consists of payment components for both regulation capacity and regulation mileage. Capacity price is paid to regulating units for reserving each MW of their generation capability in order to enable regulation service provision. Mileage price is paid for each MW of deployed regulation service (for both up-regulation and down-regulation). The payment component for regulation mileage is affected by the units' accuracy in following AGC signals. The regulation market model used in this work is based on \cite{ref26}. It is assumed that the BESS can perfectly follow AGC signals.\par

In regulation market operations, AGC signals are continuously sent to regulation market participants in short time intervals (every four seconds). Meanwhile, the amount of awarded regulation capacity and mileage for each market participant within each time interval affects the dispatch of AGC signals. Therefore, it is not trivial to model AGC signals in a bi-level optimization framework. In order to handle this issue, this paper assumes the AGC signals to have similar features as the RegD signals in the regulation market of PJM Interconnection. In this way, the average of the AGC signal is zero in each 15-minute time interval\cite{ref27}. Using this AGC signal model, when the regulation market is cleared every 15 minutes, BESS's energy level at the beginning and end of the 15-minute interval remain unchanged, if the BESS only participates in the regulation market.

This paper assumes all the three markets are cleared every 15 minutes to coordinate BESSs' resource allocation across RT energy, reserve, and regulation markets. Although this assumption increases computational efforts for solving the bi-level optimization problem, it offers two advantages. First, it enables us to model AGC signals in the biding problem. Second, clearing RT markets in sub-hour intervals is closer to the ISOs' practical operation routine.

\subsection{Formulation of The Upper-Level Problem}
In the UL problem, the BESS owner maximizes the revenue from participating in multiple markets. The UL problem is formulated as follows:
\begin{align}
    &\text{max} \sum_{t\in T}\Big[\pi_t^{E}(P_t^{BS}-P_t^{BD})+\pi_t^{Rs}P_t^{BRs}+\pi_t^{Rg,C}P_t^{BRg,C} \notag\\
    & \qquad\qquad+\pi_t^{Rg,M}P_t^{BRg,M} \Big] \Delta t \tag{U1}\\
    &\text{subject to:}  \notag \\ 
    & 0 \leq S_t^{bid} \leq u_t P^{Rate}; \; \forall t \in T  \tag{U2} \\
    & 0 \leq D_t^{bid} \leq (1-u_t)P^{Rate}; \; \forall t\in T\tag{U3} \\
    & 0 \leq Rs_t^{bid} \leq P^{Rate}; \; \forall t\in T\tag{U4} \\
    & 0 \leq Rg_t^{bid} \leq P^{Rate}; \; \forall t\in T\tag{U5} \\
    & P_t^{BD}-P_t^{BS}-P_t^{BRs}\geq -P^{Rate}+P_t^{BRg,C}; \; \forall t\in T\tag{U6} \\
    & P_t^{BD}-P_t^{BS}-P_t^{BRs}\leq P^{Rate}-P_t^{BRg,C}; \; \forall t\in T\tag{U7} \\
    & SOC_t=SOC^{Init}+\sum_{k=1}^{t}(P_k^{BD}-P_k^{BS})\Delta t; \; \forall t\in T\tag{U8}\\
    & SOC_t \geq SOC^{Min}+(P_t^{BRg,C}+P_t^{BRs})\Delta t; \; \forall t\in T\tag{U9}\\
    & SOC_t \leq SOC^{Max}-P_t^{BRg,C}\Delta t; \; \forall t\in T\tag{U10}\\
    & u_t \in \{0,1\}; \; \forall t\in T\tag{U11}
\end{align}

In the above formulation, $t$ denotes the index of market clearing intervals; $T$ denotes the timespan of the entire optimization; ${\Delta}t$ denotes the timespan of each market clearing interval; $P_t^{BS}$ and $P_t^{BD}$ denote scheduled BESS power supply and demand in RT energy market at interval $t$, respectively; $P_t^{BRs}$ denotes BESS reserve capacity at interval $t$; $P_t^{BRg,C}$ and $P_t^{BRg,M}$ denote BESS regulation capacity and regulation mileage at interval $t$, respectively; $\pi_t^E$, $\pi_t^{R_s}$, $\pi_t^{Rg,C}$, and $\pi_t^{Rg,M}$ denote the prices for BESS energy arbitrage, reserve capacity, regulation capacity, and regulation mileage at interval $t$, respectively; $S_t^{bid}$ and $D_t^{Bid}$ denote BESS supply and demand quantity bids to the RT energy market at interval $t$, respectively; $Rs_t^{bid}$ and $Rg_t^{bid}$ denote BESS reserve and regulation quantity bids at interval $t$, respectively; $P^{Rate}$ denotes BESS charging/discharging rate; $SOC_t$ denotes BESS state of charge (SOC) at interval $t$;  $SOC^{Min}$ and $SOC^{Max}$ denote BESS minimum and maximum charge levels, respectively; $SOC^{Init}$ denotes BESS initial charge level; $u_t$ is the BESS charge/discharge indicator. $u_t = 0$ or $u_t = 1$ indicates the BESS is charged or discharged during interval $t$, respectively.

In the UL problem, the objective function (U1) determines BESS's total revenue by providing different services. The constraints (U2 through U11) describes the following operating limits for BESS: 1) BESS's individual bids and total charging/discharging power at each market clearing interval lie within the charging/discharging rate of BESS; 2) For each interval, BESS is allowed to submit either a supply bid or a demand bid to the RT energy market; 3) BESS's SOC is updated at each interval based on the initial SOC and BESS's bids for energy arbitrage; 4) At each time interval, BESS's SOC lies within its upper/lower limits subtracted by the total power reserved for ancillary service provision. In practice, batteries have SOC limitations so that they will not be charged to the maximum capacity and discharged to zero. These constraints, along with batteries' charge/discharge efficiency, are neglected in this work for simplicity.

\subsection{Formulation of The Lower-Level Problem}
The LL problem describes the joint market clearing process for ISO's RT energy, reserve, and regulation markets. This LL problem is formulated as follows:
\begin{align}
    &\text{min} \sum_{t\in T}\Big(\Big[\sum_{j\in G} \big( \alpha_{j,t}^{S}P_{j,t}^{GS}+\alpha_{j,t}^{Rs}P_{j,t}^{GRs}+\alpha_{j,t}^{Rg,C}P_{j,t}^{GRg,C}+\notag\\
    & \quad \alpha_{j,t}^{Rg,M}P_{j,t}^{GRg,M} \big) \Big]+\Big[\beta_{t}^{S}P_{t}^{BS}-\beta_{t}^{D}P_{t}^{BD}+\beta_{t}^{Rs}P_{t}^{BRs}\notag\\
    &\qquad +\beta_{t}^{Rg,C}P_{t}^{BRg,C}+\beta_{t}^{Rg,M}P_{t}^{BRg,M}\Big]\Big) \Delta t \tag{L1}\\
    &\text{subject to:}  \notag \\ 
    & P_{j,t}^{GS}-P_{j,t}^{GRg,C} \geq P_j^{Min}; \; \forall t \in T,\forall j \in G  \tag{L2} \\
    & P_{j,t}^{GS}+P_{j,t}^{GRs}+P_{j,t}^{GRg,C} \leq P_j^{Max}; \; \forall t \in T,\forall j \in G  \tag{L3} \\
    & 0 \leq P_{j,t}^{GRs} \leq P^{Rs,ramp}; \; \forall t \in T,\forall j \in G\tag{L4} \\
    & 0 \leq P_{j,t}^{GRg,C} \leq P^{Rg,ramp}; \; \forall t \in T,\forall j \in G\tag{L5} \\
    & P_{j,t}^{GRg,M} \geq P_{j,t}^{GRg,C}; \; \forall t \in T,\forall j \in G\tag{L6} \\
    & P_{j,t}^{GRg,M} \leq m_j^{GRg}P_{j,t}^{GRg,C}; \; \forall t \in T,\forall j \in G\tag{L7} \\
    & 0 \leq P_{t}^{BS} \leq S_t^{bid}; \; \forall t\in T\tag{L8} \\
    & 0 \leq P_{t}^{BD} \leq D_t^{bid}; \; \forall t\in T\tag{L9} \\
    & 0 \leq P_{t}^{BRs} \leq Rs_t^{bid}; \; \forall t\in T\tag{L10} \\
    & 0 \leq P_{t}^{BRg,C} \leq Rg_t^{bid}; \; \forall t\in T\tag{L11} \\
    & P_{t}^{BRg,M} \geq P_{t}^{BRg,C}; \; \forall t \in T\tag{L12} \\
    & P_{t}^{BRg,M} \leq m^{BRg}P_{t}^{BRg,C}; \; \forall t \in T\tag{L13} \\
    & \sum_{j\in G}P_{j,t}^{GRs} + P_t^{BRs} \geq Q_t^{Rs}; \; \forall t\in T : \; \pi_t^{Rs}\tag{L14}\\
    & \sum_{j\in G}P_{j,t}^{GRg,C} + P_t^{BRg,C} \geq Q_t^{Rg,C}; \; \forall t\in T : \; \pi_t^{Rg,C}\tag{L15}\\
    & \sum_{j\in G}P_{j,t}^{GRg,M} + P_t^{BRg,M} \geq Q_t^{Rg,M}; \; \forall t\in T : \; \pi_t^{Rg,M}\tag{L16}\\
    & \sum_{j\in G}P_{j,t}^{GS} + P_t^{BS}-P_t^{BD} = P_t^{Load}; \; \forall t\in T : \; \pi_t^{E}\tag{L17}
\end{align}

In the above formulation, $G$ denotes the set of regular generating units (other than BESS); $\alpha_{j,t}^{S}$, $\alpha_{j,t}^{Rs}$, $\alpha_{j,t}^{Rg,C}$, and $\alpha_{j,t}^{Rg,M}$ denote the energy, reserve, regulation capacity, and regulation mileage price bids for $j^{th}$ generating unit at interval $t$, respectively; $P_{j,t}^{GS}$, $P_{j,t}^{GRs}$, $P_{j,t}^{GRg,C}$, and $P_{j,t}^{GRg,M}$ denote the scheduled power of $j^{th}$ generating unit at interval $t$, for energy, reserve, regulation capacity, and regulation mileage provision, respectively; $\beta_t^{S}$, $\beta_t^{D}$, $\beta_t^{Rs}$, $\beta_t^{Rg,C}$, and $\beta_t^{Rg,M}$ denote BESS's price bids for energy supply, energy demand, reserve, regulation capacity and regulation mileage provision at interval $t$, respectively; $P_j^{Min}$ and $P_j^{Max}$ denote the minimum and maximum output of $j^{th}$ generating unit, respectively; $P^{Rs,ramp}$ and $P^{Rg,ramp}$ denote the reserve and regulation ramp rates of the generating unit, respectively; $Q_t^{Rs}$, $Q_t^{Rg,C}$ and $Q_t^{Rg,M}$ denote the amount of system reserve, regulation capacity and regulation mileage requirements at interval $t$, respectively; $P_t^{Load}$ denotes system total load at interval $t$; $m_j^{GRg}$ and $m^{BRg}$ denote regulation mileage multipliers of $j^{th}$ generating unit and the BESS, respectively. These multipliers are calculated by ISOs based on the market participants' historical performance on regulation service provision.

In the LL problem, the objective function (L1) determines the total operating cost of the system considering energy, reserve, and regulation market operations. The constraints (L2 through L17) describes the following operating limits for ISO markets: 1) For each generating unit, its total power delivery at each time interval lies within its maximum and minimum generation limits; 2) For each generating unit, its reserve and regulation capacity provision at each time interval does not exceed the corresponding ramp rates; 3) Each regulation market participant (including the BESS and generating units) satisfies the regulation market requirements defined in (L6), (L7), (L12) and (L13); 3) The scheduled power of BESS in different markets is limited by the corresponding power bids; 4) System requirements for reserve, regulation capacity, and regulation mileage provision are satisfied when the markets are cleared at each time interval; 5) The system total load needs to be served at each time interval. For simplicity, the above LL problem does not model the transmission system.

\subsection{Solution Procedure}
The proposed bi-level optimization is a nonlinear and non-convex problem. For solving this problem, it is converted to a MILP problem. Details of this conversion process can be found in \cite{ref28}.

\section{Case Studies}
In this section, numerical studies are performed to evaluate the performance of the proposed framework, investigate the behavior of a price-maker BESS across multiple markets, and study the BESS's impact on various markets.

\subsection{The Test System}
The test system contains five generators and one utility-scale BESS. Numerical studies are performed over a 24-hour horizon consisting of 96 market clearing intervals. The length of each market clearing interval is 15 minutes. Table \ref{T1} shows the generators' parameters. The BESS has an energy capacity of 400 MWh and a charging/discharging rate of 40 MW. 

\begin{figure}[!t]
	\centering
	\includegraphics[width=3.1in]{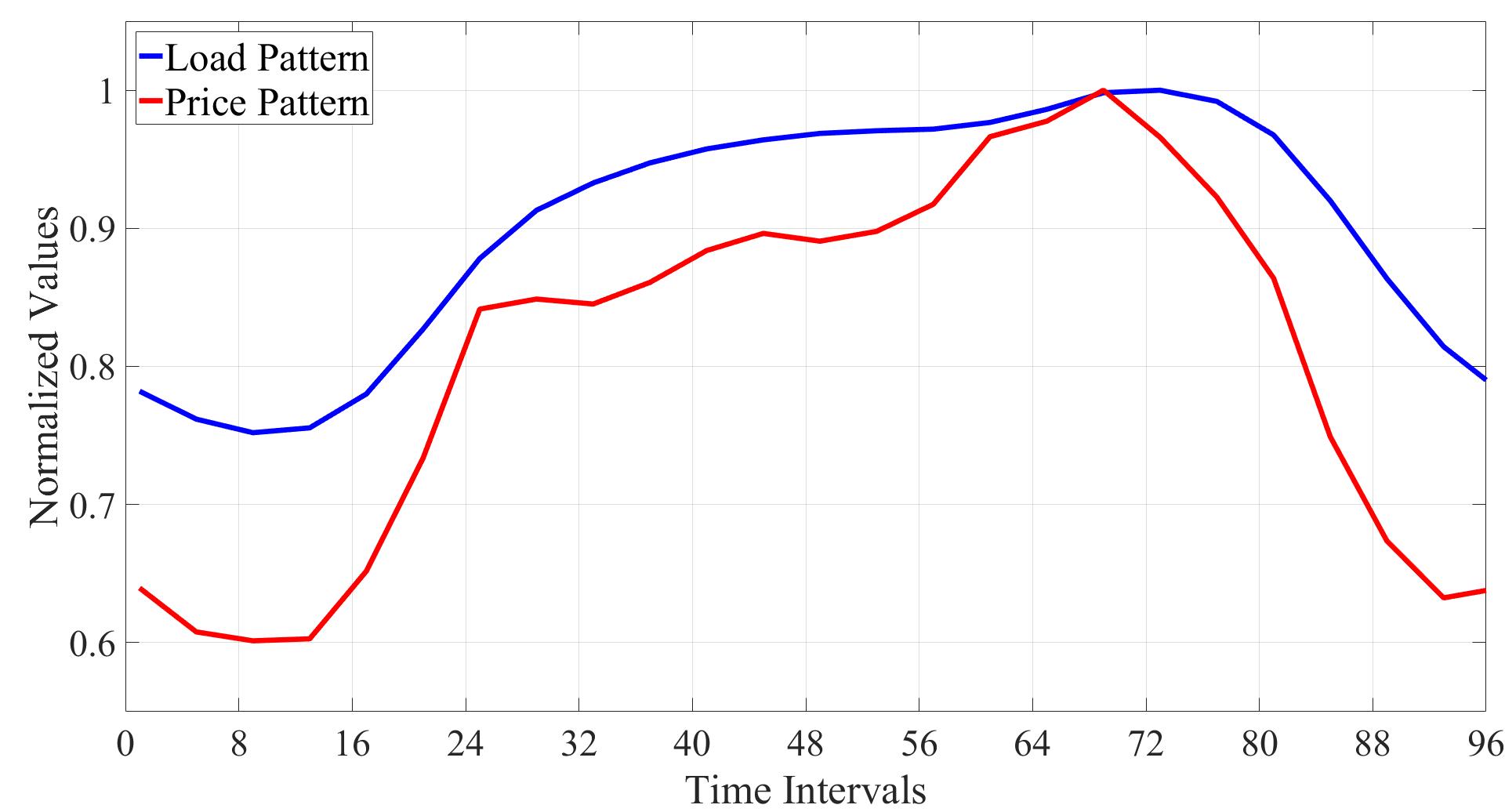}
	\caption{Normalized price and load patterns used in the case studies.}
	\label{pattern}
\end{figure}

The numerical case studies require input data for the system total loads and the price bids of all the other units. In this paper, these inputs are created based on the historical price and load data from PJM Interconnection \cite{ref29}. Fig. \ref{pattern} shows the normalized price and load patterns over the 24-hour horizon (extracted from PJM 2018 data). 

Total system demand of 1000 MW is mapped on the load pattern for creating the system load in each time interval of simulations. It means that, according to Fig. \ref{pattern}, system load is 1000 MW in the $73^{th}$ interval and is less than that in other intervals. System's spinning reserve and regulation capacity requirement are considered to be 10\% and 4\% of load in each interval. Regulation mileage requirement is assumed to be 1.75 times the regulation capacity requirement in each interval.

The base price bid of each generator, which is given in Table 1 is also mapped on the extracted price pattern to create each generator's energy price bid for each time interval. Additionally, in 2018 PJM data, average ratios of reserve price, regulation capacity price, and regulation mileage price to the energy price are respectively 0.15, 0.4, and 0.07. Therefore, generators' energy price bid in each time interval is multiplied by 0.15, 0.4 and 0.07 to respectively create their price bids for providing the reserve, regulation capacity, and regulation mileage. Note that the mileage multipliers of generators and BESS are assumed large enough to not to limit their outputs.

\begin{table}[!t]
	\caption{Generator Information}
	\label{T1}
	%\centering
	\begin{tabular}{c||cccc}
		\hline
		Generator & Base Price Bid & $P^{Max}$ & $P^{Rs,ramp}$ & $P^{Rg,ramp}$\\
		Number & (\$/MWh) & (MW) & (MW) & (MW) \\
		\hline
		\hline
		1 & 10 & 400 & 80 & 40 \\
		2 & 14 & 300 & 60 & 30 \\
		3 & 15 & 210 & 42 & 21 \\
		4 & 30 & 350 & 70 & 35 \\
		5 & 40 & 270 & 54 & 27 \\
		\hline
	\end{tabular}
\end{table}

\subsection{Case Study Results}
The proposed approach is tested in four different cases that represent various market participation policies for the BESS.

\begin{figure}[!t]
	\centering
	\includegraphics[width=3.1in]{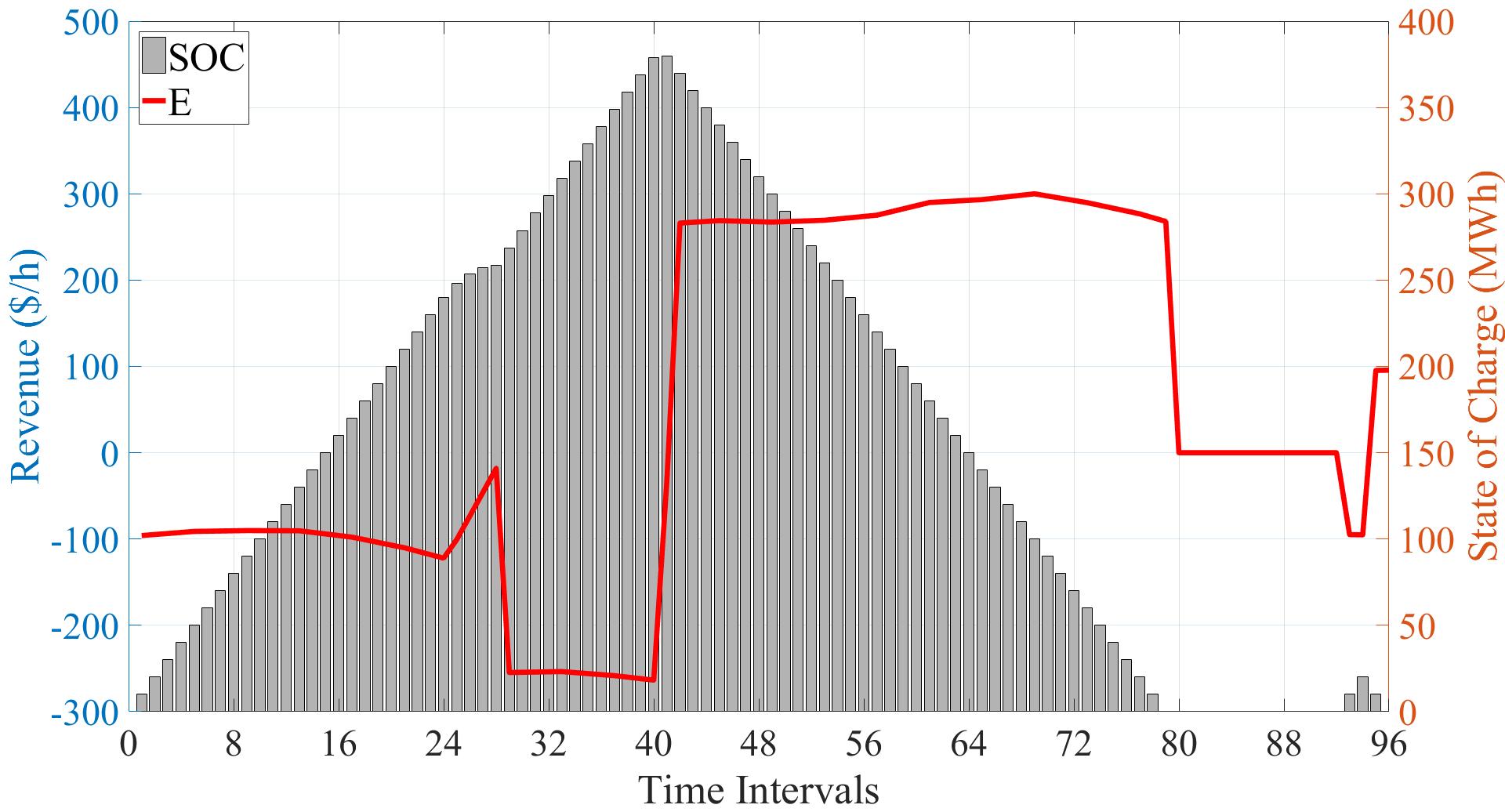}
	\caption{Simulation result of Case 1. SOC indicates the state of charge of the BESS at each interval; E indicates BESS revenue from the energy market.}
	\label{E}
\end{figure}

In Case 1, only the energy market is modeled, and BESS performs energy arbitrage between different hours. Fig. \ref{E} presents the revenue values and SOC of the BESS across the 24-hour horizon. It is shown in Fig. \ref{E} that the BESS charges during off-peak hours when the energy price is low and discharges during peak hours when the energy price is high. Negative revenue from the energy market indicates the BESS's energy purchasing activities through the RT energy market. It can be seen that in the last intervals, price changes in a way that it is profitable for BESS to buy and then sell energy to the market, which also causes changes in SOC. Note that ancillary service markets are not considered in this case.

\begin{figure}[!t]
	\centering
	\includegraphics[width=3.1in]{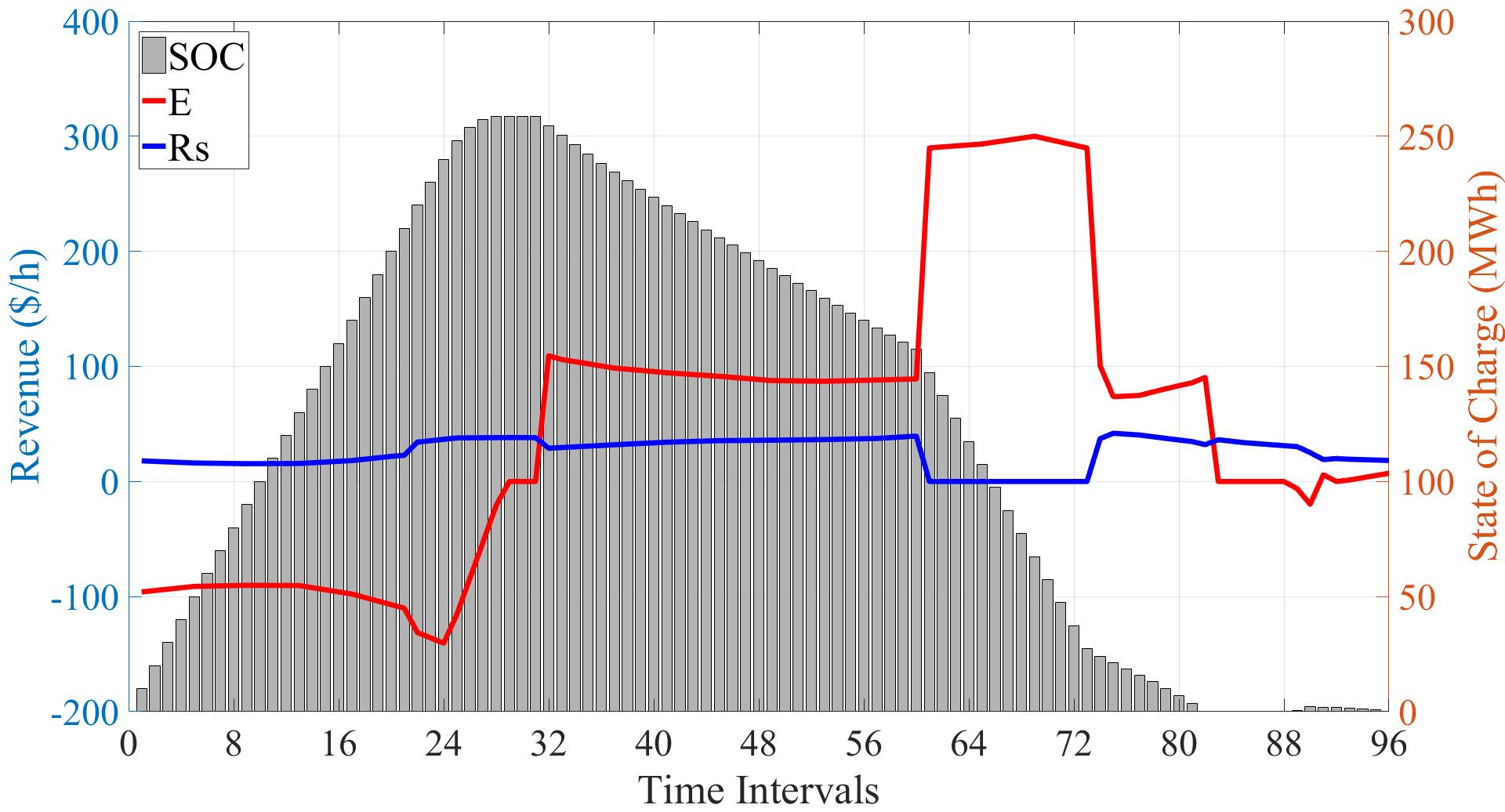}
	\caption{Simulation result of Case 2. SOC indicates the state of charge of the BESS at each interval; E indicates BESS revenue from the energy market; Rs indicates BESS revenue from the spinning reserve market.}
	\label{ERs}
\end{figure}

In Case 2, energy and spinning reserve markets are modeled. Fig. \ref{ERs} presents BESS's 24-hour revenue and SOC for this case. According to Fig. \ref{ERs}, when the BESS participates in both energy and reserve markets, it first charges itself during the off-peak hours at the beginning of the day, when the energy price is low. During the peak hours with high energy prices, the BESS allocates its available energy optimally across both energy and reserve markets. It is observed that during the charging period, BESS has negative revenue from the energy market and positive revenue from the reserve market, when its SOC increases with a constant rate. It means that the BESS simultaneously purchases energy from the energy market and sells all or part of its purchased energy to the reserve market. This represents the BESS's arbitrage activities between these two markets at the same market clearing interval. This arbitrage between energy and reserve markets could not happen during the discharging period (when the SOC decreases), due to the BESS operating constraints (U6 and U7) defined in the UL problem. In other words, constraints U6 and U7 indicate that 1) during charging periods, BESS could provide reserve service by reducing the amount of real power it purchases from the real-time energy market; 2) during discharging periods, BESS could not perform the above activity, since it acts as a seller (instead of a buyer) in the real-time energy market and it needs extra capacity for reserve service provision.

\begin{figure}[!t]
	\centering
	\includegraphics[width=3.1in]{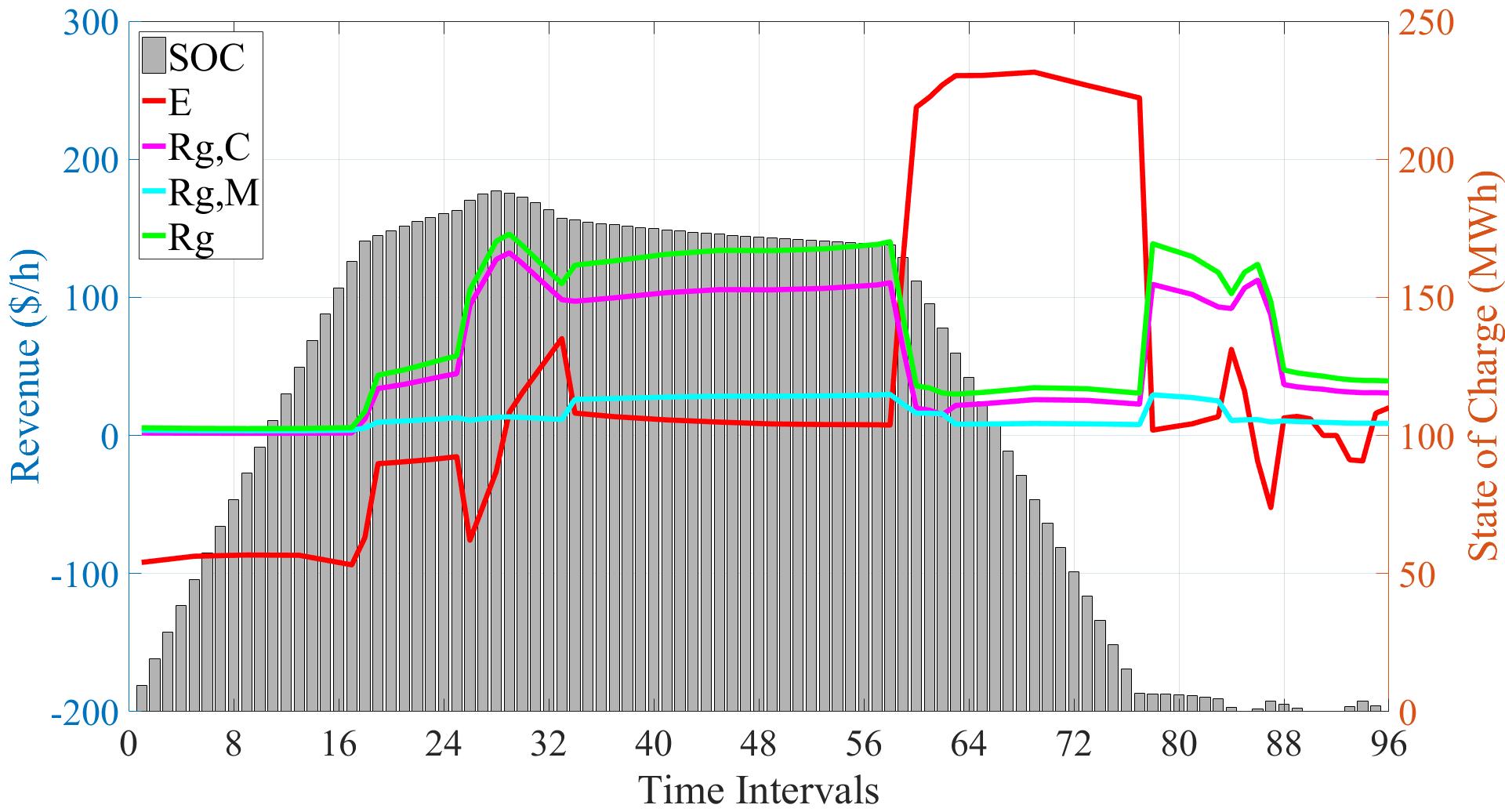}
	\caption{Simulation result of Case 3. SOC indicates the state of charge of the BESS at each interval; E indicates BESS's revenue from the energy market; Rg,C indicates BESS's revenue from regulation capacity provision; Rg,M indicates BESS's revenue from regulation mileage provision; Rg indicates BESS's total revenue from the regulation market (considering both regulation capacity and mileage provision).}
	\label{ERg}
\end{figure}

In Case 3, energy and regulation markets are modeled. Fig. \ref{ERg} shows BESS's 24-hour revenue and SOC under this scenario. According to Fig. \ref{ERg}, during the charging period (when the SOC increases), the BESS does not perform arbitrage activities between the energy and regulation markets at the same market clearing interval. This is caused by the fact that in order to participate in the regulation market, the BESS's SOC needs to first reach a sufficient level for providing regulation services in both directions. These operating constraints (U6 and U7) are defined in the UL problem. To meet these constraints, during the off-peak hours, the BESS first charges itself to reach a profitable SOC level. After the initial charging period, the BESS allocates its available energy to participate in both energy and regulation markets. Comparison of Fig. \ref{ERg} and Fig. \ref{ERs} shows that when the BESS participates in both energy and regulation markets, its revenue from the energy market during the peak hours is lower compared to Case 2. On the other hand, in Case 3, similar to Case 2, the maximum SOC reached by the BESS across simulation horizon is reduced compared to Case 1. This indicates that BESS could avoid deep charging/discharging cycles by participating in ancillary services markets, which can reduce batteries' degradation cost.

\begin{figure}[!t]
	\centering
	\includegraphics[width=3.1in]{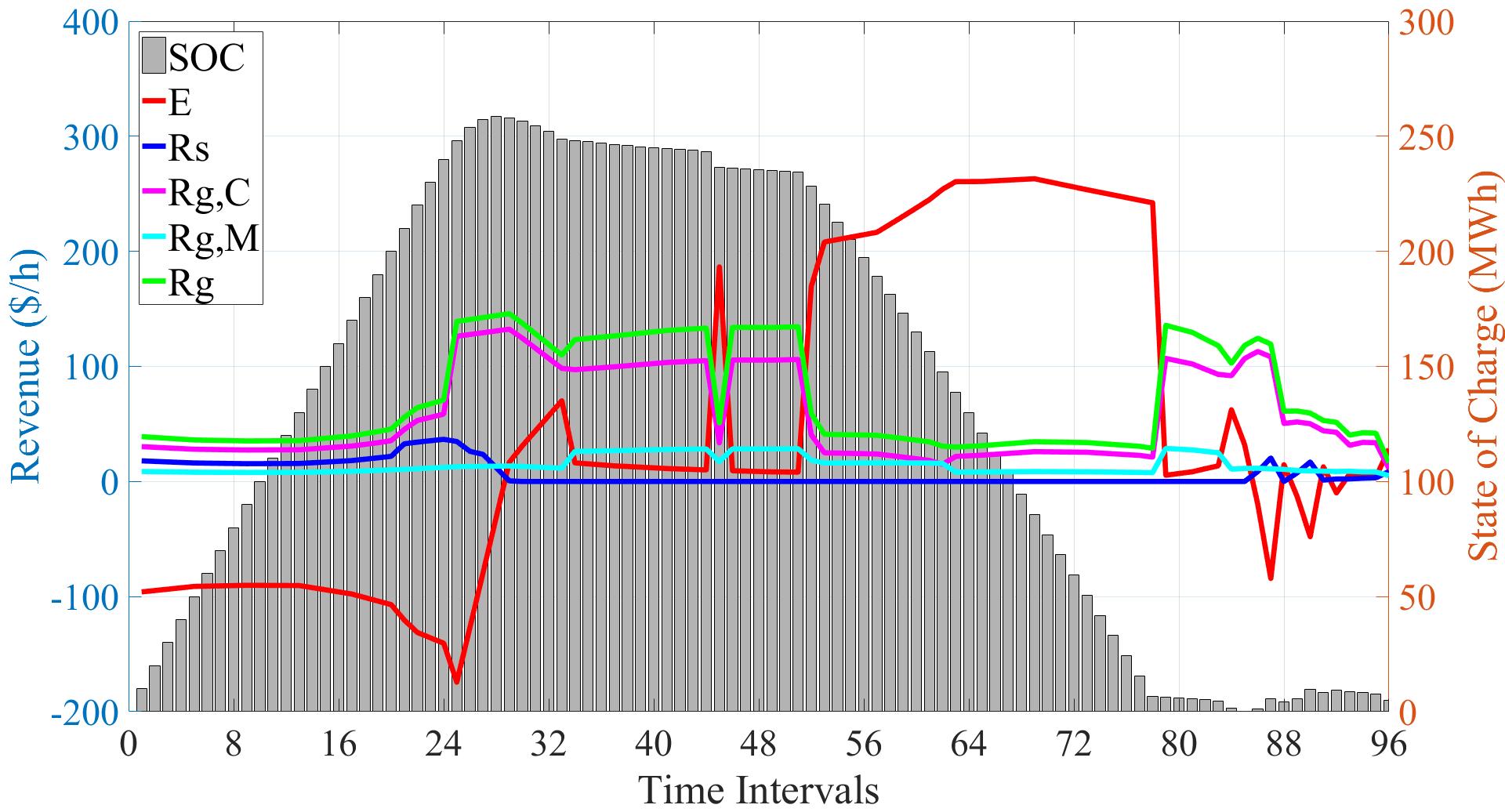}
	\caption{Simulation result of Case 4. SOC indicates the state of charge of the BESS at each interval; E indicates BESS's revenue from the energy market; Rs indicates BESS revenue from the spinning reserve market; Rg,C indicates BESS's revenue from regulation capacity provision; Rg,M indicates BESS's revenue from regulation mileage provision; Rg indicates BESS's total revenue from the regulation market (considering both regulation capacity and mileage provision).}
	\label{ERsRg}
\end{figure}

In Case 4, all of the energy, reserve, and regulation markets are modeled together. Fig. \ref{ERsRg} shows BESS's 24-hour revenue and SOC under this scenario. It is shown in Fig. \ref{ERsRg} that the BESS obtains the least amount of revenue from the reserve market. The BESS only performs limited arbitrage activities between the energy and reserve markets at the charging period. During the discharging period, the BESS do not participate in the reserve market for most of the time. Additionally, by comparing BESS's revenue patterns around the $45^{th}$ time interval between Case 3 and Case 4, one could observe that: 1) in both cases, the BESS does not gain revenue from the reserve market; 2) BESS's revenue from energy and regulation markets in Case 4 is significantly different from that in Case 3. This indicates the BESS could affect market outcomes in multiple ways once it has the option of participating in all the energy and ancillary services markets.

\begin{figure}[!t]
    \centering
    \includegraphics[width=3.1in]{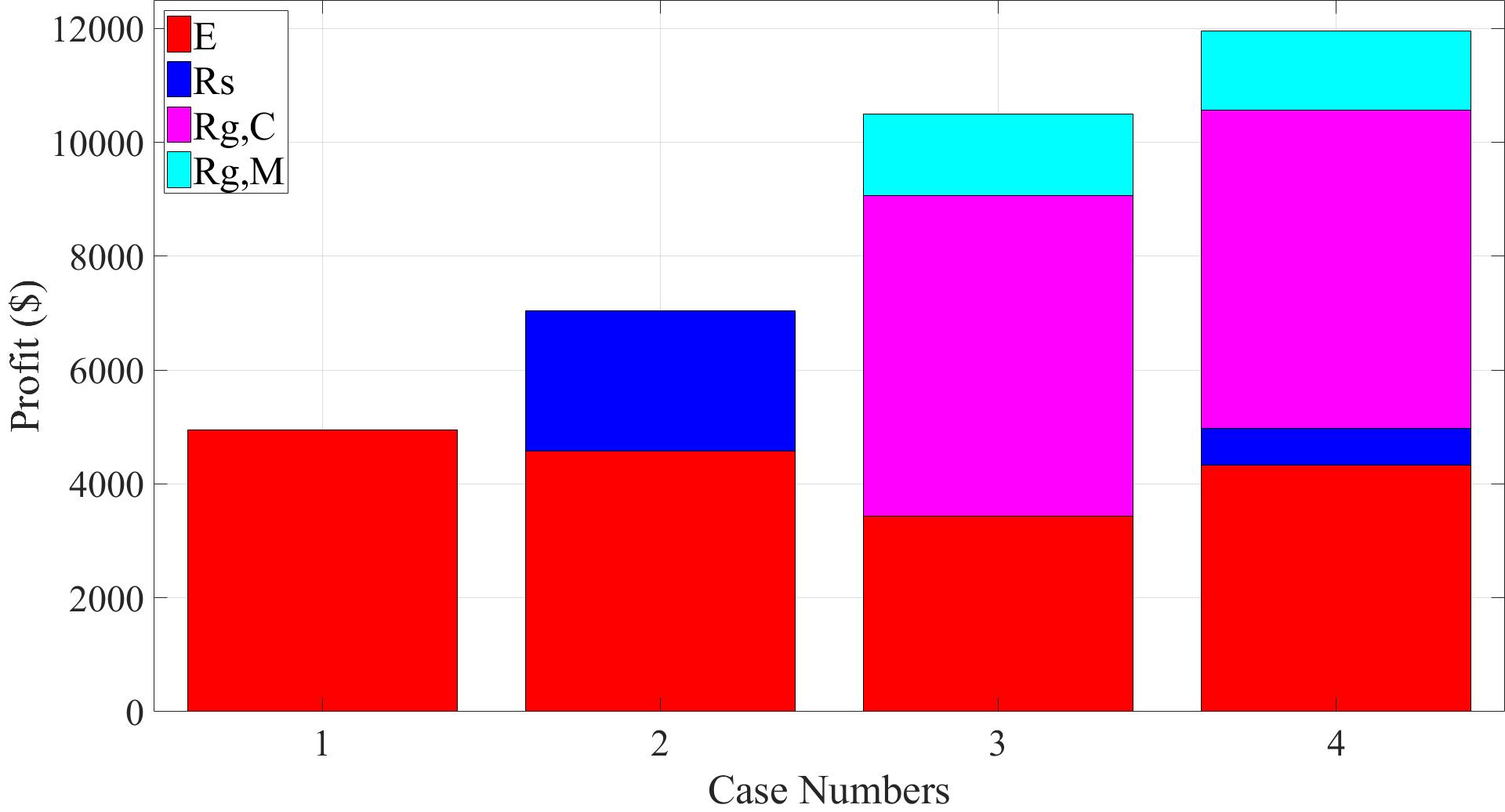}
    \caption{BESS profit in various cases. E indicate BESS's revenue from the energy market; Rs indicate BESS's revenue from the spinning reserve market; Rg,C indicate BESS's revenue from regulation capacity provision; Rg,M indicate BESS's revenue from regulation mileage provision.}
    \label{R}
\end{figure}

Fig. \ref{R} shows BESS's total revenue from various markets for Case 1 through Case 4. Fig. \ref{R} indicates: 1) the BESS tends to participate more in the regulation market and obtain most of its revenue from it; 2) the BESS could gain more profit once it is allowed to participate in more markets. These observations agree with the BESS operating patterns in real-world practices \cite{ref100}. Besides, by comparing BESS's revenue from the energy market in Case 3 and Case 4, it is observed that the BESS could obtain more revenue from the energy market if it is allowed to participate in the reserve market.

\section{Conclusion}
This paper presents a bi-level optimization framework to study the strategic bidding problem of a price-maker BESS across energy, spinning reserve, and pay as performance regulation markets. A proper approach is applied to model the AGC signals in this bi-level framework. A realistic way of generating synthetic test case data is applied to map the historical price and load patterns in PJM Interconnection to the studied test system. Case study results show the impact of a price-maker BESS on the joint market clearing process.

Built upon this work, future research could focus on considering the BESS degradation cost and refining the AGC signal modeling in the proposed bi-level framework, as well as studying the impact of the transmission network on the operations of BESS-integrated wholesale electricity markets.

\bibliographystyle{IEEEtran}
\bibliography{IEEEabrv,Ref}



% \end{thebibliography}

\end{document}